# Angular momentum effects in neutron decay

I. Pavlov ®,* A. Chaikovskaia ®,† and D. Karlovets ®‡
*School of Physics and Engineering, ITMO University, 197101 St. Petersburg, Russia*

We investigate the intriguing phenomenon of beta decay of a free neutron in a non-plane-wave (structured) state. Our analysis covers three types of states: unpolarized vortex (Bessel) neutrons that possess nonzero orbital angular momentum (OAM), Laguerre-Gaussian wave packets, and spin-correlated OAM (spin-orbit) states characterized by unique polarization patterns. These states are of particular interest as they have recently been generated in neutron optics experiments and have promising applications in studies of quantum magnetic materials. The spectral-angular distributions (SAD) of the emitted electrons and protons are examined. We show that the high sensitivity of the protons' SAD to the structure of the neutron wave packet can be used as a tool to extract the distinctive features of the non-plane-wave neutron states. Furthermore, we demonstrate that the angular distribution of the emitted particles serves as a reflection of the spatial symmetries inherent to the neutron wave packet.

## I. INTRODUCTION

Neutrons play a crucial role in the structure of atomic nuclei and are essential for understanding the fundamental forces that govern matter. Their unique properties and interactions make them vital for both theoretical studies in particle physics and practical applications [1]. Recent advancements in neutron optics have enabled the generation and detection of neutrons in structured quantum states, which go beyond simple plane waves [2,3]. Among these are so-called vortex states with a phase singularity and a corresponding orbital angular momentum (OAM) projection onto a propagation direction [4–6], spin correlated OAM (spin-orbit) states [7–9], and Airy beams [10], which have emerged as promising tools for material characterization and fundamental physics exploration.

In recent years, there has been a surge of theoretical interest in various phenomena involving vortex neutrons, such as elastic scattering by nuclei [11,12] and hydrogen targets [13] and radiative capture of neutrons by protons [14]. Most notably, the decay of a twisted neutron was recently investigated using a simple model of an unpolarized Bessel vortex state [15]. This study revealed significant differences in the decay patterns of vortex-state neutrons compared to plane-wave ones, particularly in the angular and energy distributions of the emitted particles. In this paper, we delve deeper into the decay of vortex neutrons, focusing on both Bessel states and Laguerre-Gaussian vortex neutron beams [7,16]. We also investigate the implications of the neutron's polarization in the context of spin-orbit coupled states [7,8], with a particular interest in the spectral-angular distributions of emitted protons.

While the decay of spin-orbit neutron states remains uncharted territory, weak decays of other particles in polarized non-plane-wave states have been studied, such as the decay of vortex muons [17,18]. These investigations have shown that the final electron spectra can be significantly altered compared to those produced by plane-wave muons. Moreover, the angular distributions of emitted particles often reflect the spatial symmetries inherent to their initial polarization states [18].

The structure of this paper is as follows. In Sec. II, we explore the kinematics of plane-wave neutron decay in the laboratory frame. Sec. III presents our calculations of the decay rate for a plane-wave polarized neutron. Building on this foundation, we extend our findings to arbitrary initial neutron states and discuss specific cases: Bessel vortex states, Laguerre-Gaussian wave packets and spin-orbit states. In Sec. IV, we provide and analyze our numerical results. Finally, we conclude with a summary of our findings and their implications.

Throughout the paper, the natural system of units $\hbar = c = 1$ is used.

## II. KINEMATICS

The beta decay of a free neutron is a process given by $n(p_n) \to e^-(p_e) + \bar{\nu}_e(p_{\bar{\nu}_e}) + p(p_p)$. Typically, the decay is analyzed within the neutron rest frame [19–21]. This approach is practical for experiments involving cold and ultracold neutrons [22,23], where their velocities can be considered negligible. Since the structured neutron state can be thought of as a superposition of plane waves propagating in different directions, a common rest frame for them does not exist. Therefore, an essential preliminary step in studying the decay of a structured neutron wave packet is the analysis of a plane-wave neutron decay with nonzero momentum in the laboratory frame, even if the neutron is low energetic.

---

*Contact author: ilya.pavlov@metalab.ifmo.ru
†Contact author: alisa.katanaeva@metalab.ifmo.ru
‡Contact author: dmitry.karlovets@metalab.ifmo.ru

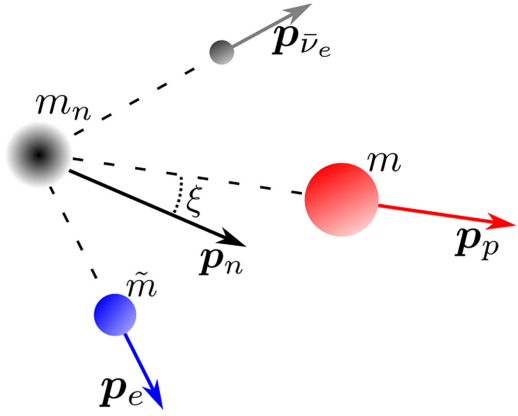

FIG. 1. Kinematics of decay.

Let us consider a neutron with mass $m_n$, momentum $\bm{p}_n$ and energy $E_n = \sqrt{\bm{p}_n^2 + m_n^2}$. Suppose that one of the massive particles emitted—for example proton with mass $m$—is detected (as a plane wave) at the deflection angle $\xi$ with respect to the neutron propagation axis (see Fig. 1). The energy of the detected particle in the laboratory frame is denoted as $E$. The allowed energy ranges for the final particles are easier to determine in the neutron rest frame; that is why we connect the energies in two frames via the corresponding Lorentz boost:

$$\gamma(E - \beta \cos\xi \sqrt{E^2 - m^2}) = e, \tag{1}$$

where $e$ is the energy in the neutron rest frame, $\beta = |\bm{p}_n|/E_n$, and $\gamma = 1/\sqrt{1-\beta^2}$. This relation can be regarded as the equation for $E$, with solutions being

$$E(e) = \frac{e \pm \beta \cos\xi \sqrt{e^2 - \gamma^2(1-\beta^2\cos^2\xi)m^2}}{\gamma(1-\beta^2\cos^2\xi)} \tag{2}$$

with the condition $E > m$. The minimal energy of the particle in the neutron rest frame obviously equals $m$ while the maximal energy in this frame can be found by using the conservation laws [19]:

$$e_{\max} = \frac{m_n^2 - \tilde{m}^2 + m^2}{2m_n}, \tag{3}$$

where $\tilde{m}$ is the mass of another emitted massive particle, in this case the electron.

One should, however, be very cautious with determining the allowed range of $E$ by using Eq. (2), because minimal and maximal values of $E$ do not necessarily correspond to those of $e$.

In order to properly describe the energy range one should introduce the "critical" neutron velocity

$$\beta_{\mathrm{cr}} = \frac{\sqrt{[(m_n+\tilde{m})^2 - m^2][(m_n-\tilde{m})^2 - m^2]}}{m^2 + m_n^2 - \tilde{m}^2}. \tag{4}$$

For $\beta \geqslant \beta_{\mathrm{cr}}$ there exists a maximal deflection angle $0 \leqslant \xi_{\max} \leqslant \pi/2$, such that

$$\cos\xi_{\max}(\beta) = \frac{1}{\beta}\sqrt{1 - \frac{(1-\beta^2)(m_n^2 - \tilde{m}^2 + m^2)}{4m^2 m_n^2}}. \tag{5}$$

For angles larger than $\xi_{\max}$ the decay is not allowed as the expression under the square root in Eq. (2) becomes negative even for maximal energy:

$$e_{\max}^2 - \gamma^2(1-\beta^2\cos^2\xi)m^2 < 0. \tag{6}$$

The critical angle equals $\frac{\pi}{2}$ when $\beta = \beta_{\mathrm{cr}}$. For $\beta \geqslant \beta_{\mathrm{cr}}$ and $\xi < \xi_{\max}$ the decay is allowed; maximal and minimal energy are given by Eq. (2) with $e = e_{\max}$ substituted and with the "+" and "−" signs, respectively:

$$E_{\max,\min} = \frac{m_n^2 - \tilde{m}^2 + m^2 \pm \beta\cos\xi\sqrt{(m_n^2 - \tilde{m}^2 + m^2)^2 - 4\gamma^2(1-\beta^2\cos^2\xi)m^2 m_n^2}}{2m_n\gamma(1-\beta^2\cos^2\xi)}. \tag{7}$$

Importantly, the minimal energy here is *not* equal to the rest energy $m$, which was not reflected in a recent work [15]. Note that for protons $\beta_{\mathrm{cr}} \approx 0.00126$ (neutron's kinetic energy $\approx 750\,\mathrm{eV}$) while for electrons $\beta_{\mathrm{cr}} \approx 0.9181$ (kinetic energy 1.4 GeV). Thus, the effect of the critical angle can hardly be observed in the electron angular distributions due to the experimentally feasible range of neutron's velocity $\beta$ [24], whereas for the proton distributions it seems more promising.

For "slow" neutrons ($\beta < \beta_{\mathrm{cr}}$) the critical angle does not exist. The minimal energy of the final particle is simply equal to its rest energy $m$ while the maximal energy is provided by the same expression as before:

$$E_{\max} = \frac{m_n^2 - \tilde{m}^2 + m^2 + \beta\cos\xi\sqrt{(m_n^2 - \tilde{m}^2 + m^2)^2 - 4\gamma^2(1-\beta^2\cos^2\xi)m^2 m_n^2}}{2m_n\gamma(1-\beta^2\cos^2\xi)}, \tag{8}$$

$$E_{\min} = m. \tag{9}$$

## III. DECAY RATE

### A. Plane wave neutron in the laboratory frame

The standard model calculation of the plane wave decay amplitude at tree-level yields

$$\mathcal{M} = \frac{G_F}{\sqrt{2}} [\bar{u}(p_p)\gamma^\mu (1-\gamma^5) u(p_n)][\bar{u}(p_e)\gamma_\mu (1-\gamma^5) v(p_{\bar{\nu}_e})], \tag{10}$$

where $G_F \approx 1.166 \times 10^{-5}$ GeV$^{-2}$ is the Fermi constant. Here we limit ourselves to the Fermi four-fermion interaction theory and for simplicity neglect the vector and axial vector form factors [19]. The plane wave $S$-matrix element is connected with the decay amplitude via the relation

$$\mathcal{S}_{\text{PW}}(\mathbf{p}_n) = i(2\pi)^4 \delta^{(4)}(p_n - p_p - p_e - p_{\bar{\nu}_e}) \cdot \mathcal{M}, \tag{11}$$

and the decay rate is given by the Fermi golden rule

$$d\Gamma_{\text{PW}} = \frac{|\mathcal{M}|^2}{2E_n} \left(\frac{d^3 p_e}{(2\pi)^3 2E_e}\right)\left(\frac{d^3 p_p}{(2\pi)^3 2E_p}\right)\left(\frac{d^3 p_{\bar{\nu}_e}}{(2\pi)^3 2E_\nu}\right)$$
$$\times (2\pi)^4 \delta^4(p_n - p_p - p_e - p_{\bar{\nu}_e}). \tag{12}$$

For the unpolarized neutron we average over its polarizations and sum over the polarizations of all final particles. By using the well-known relations

$$\sum_s u(p)\bar{u}(p) = \slashed{p} + m, \quad \sum_s v(p)\bar{v}(p) = \slashed{p} - m, \tag{13}$$

one obtains

$$\langle |\mathcal{M}|^2 \rangle = 64 G^2 (p_n p_{\bar{\nu}_e})(p_e p_p). \tag{14}$$

The observables that we would like to investigate in this paper are the spectral-angular distributions (SAD) of the outgoing electrons and protons, the expressions for which are mathematically equivalent up to a permutation of masses $m_e$ and $m_p$. In order to present all equations in this section in a general form, we use indices $i$ and $j$:

$$\text{if } i = e, \text{ then } j = p, \tag{15}$$

and vice versa. We assume that *only one final particle is detected*, thus the differential decay rate is to be integrated over the phase space of neutrino and the second massive particle. For these purpose, we use the technique similar to that described in [25] (see Appendix A for the details). From that calculation we obtain the decay rate

$$\frac{d\Gamma}{dE_j d\Omega_j} = \frac{G^2}{48\pi^4 E_n}\left(1 - \frac{m_i^2}{q^2}\right)\left[\left((p_n p_j)(3m_j^2 + 3m_n^2 - 4(p_n p_j)) - 2m_n^2 m_j^2\right)\left(1 - \frac{m_i^2}{2q^2} - \frac{m_i^4}{2q^4}\right)\right.$$
$$\left. - \frac{3}{2}\left(2m_n^2 m_j^2 - (p_n p_j)(m_n^2 + m_j^2)\right)\frac{m_i^2}{q^2}\left(1 - \frac{m_i^2}{q^2}\right)\right]\sqrt{E_j^2 - m_j^2}, \tag{16}$$

where

$$q^2 = (p_n - p_j)^2 = m_n^2 + m_j^2 - 2(p_n p_j), \tag{17}$$

$$(p_n p_j) = E_j \sqrt{m_n^2 + \mathbf{p}_n^2} - \mathbf{p}_n \cdot \mathbf{p}_j. \tag{18}$$

Calculations for the polarized neutron are no more complicated than for the unpolarized one. In this case there is an additional characteristic of the neutron, spin four-vector $s^\mu$, which has the following properties:

$$s^\mu p_\mu = 0; \tag{19}$$

$$s^\mu s_\mu = -1. \tag{20}$$

The last equality holds because we deal only with pure polarization states. In the neutron rest frame

$$s^\mu = (0, \mathbf{s}) \tag{21}$$

whereas in the laboratory frame

$$s^\mu = (\gamma \beta (\mathbf{s} \cdot \mathbf{n}), \ \mathbf{s} + (\gamma - 1)\mathbf{n}(\mathbf{s} \cdot \mathbf{n})), \tag{22}$$

where $\mathbf{n} = \mathbf{p}_n / |\mathbf{p}_n|$. Instead of averaging over polarizations as in Eq. (13) we use the following relation (see [25,26]):

$$u(p_n, s)\bar{u}(p_n, s) = \tfrac{1}{2}(\slashed{p}_n + m_n)(1 - \gamma_5 \slashed{s}). \tag{23}$$

One can show that it only yields the change

$$p_n^\mu \to p_n^\mu - m_n s^\mu \tag{24}$$

in the decay rate (A3). This trick is shown in [25] for the muon decay in the massless limit, but the nonzero masses of the final particles do not violate this feature.

### B. Vortex Bessel neutron

Now we proceed to the vortex Bessel neutron [5,11–13,15]. It can be described as a specific monochromatic superposition of plane waves,

$$|\kappa, \ell, p_z\rangle = \int \frac{d^2 \mathbf{p}_n}{(2\pi)^2} a_{\kappa \ell}(\mathbf{p}_n) |\text{PW}(\mathbf{p}_n)\rangle, \tag{25}$$

where we imply that each individual plane wave $|\text{PW}(\mathbf{p}_n)\rangle$ in the superposition has a definite momentum $p_z$ along the $z$ axis and contains a bispinor $u(p_n, s)$. The amplitude is given by

$$a_{\kappa \ell}(\mathbf{p}) = (-i)^\ell e^{i\ell \varphi_n} \sqrt{\frac{2\pi}{\kappa}} \delta(|\mathbf{p}_\perp| - \kappa), \tag{26}$$

where $\ell$ is the is the azimuthal quantum number indicative of OAM and $\kappa$ is the absolute value of transverse momentum. As we run through all the plane-wave components inside a Bessel state, their momentum vectors rotate around the $z$ axis and trace out the surface of a cone with an opening angle $\theta_n$, defined as

$$\cos\theta_n = p_z/|\boldsymbol{p}_n|, \quad \sin\theta_n = \kappa/|\boldsymbol{p}_n|. \tag{27}$$

Generally, the polarization vector $\boldsymbol{s}$ of each plane-wave component can be an arbitrary function of the azimuthal angle $\varphi_n$, manifesting the spin-orbit interaction [27] and resulting in inhomogeneous polarization state [17,18]. One of the simplest examples used in the literature is specifying the helicity of each plane-wave state ($\boldsymbol{s} = \boldsymbol{p}_n/|\boldsymbol{p}_n|$). Consequently, the $S$-matrix element becomes

$$\mathcal{S}_{\text{tw}} = \int \frac{d^2 \boldsymbol{p}_n}{(2\pi)^2} a_{\kappa\ell}(\boldsymbol{p}_n)\mathcal{S}_{\text{PW}}(\boldsymbol{p}_n). \tag{28}$$

In Appendix B we demonstrate that the decay width of a Bessel vortex neutron is reduced to the azimuthal average of the corresponding plane-wave decay widths [15,17,28]:

$$d\Gamma_{\text{tw}} = \int \frac{d\varphi_n}{2\pi} d\Gamma_{\text{PW}}. \tag{29}$$

The kinematic peculiarities of the Bessel neutron decay lie solely in the fact that the energetic and angular constraints for the final detected particle are different for all plane-wave components of the decaying neutron state.

The geometry of the problem is simplified if the vortex neutron is either unpolarized or its polarization state is azimuthally invariant [17]. This results in *azimuthally symmetric* distributions of the final particles. In this case we can choose the momentum of the final $j$ particle to define the $(x, z)$ plane: $\boldsymbol{n}_j = (\sin\theta_j, 0, \cos\theta_j)$. In this coordinate system, any chosen plane-wave component of the vortex state is defined by $\boldsymbol{n} = (\sin\theta_n\cos\varphi_n, \sin\theta_n\sin\varphi_n, \cos\theta_n)$. Thus, the $\cos\xi$ in all previous plane-wave expressions in Sec. II should be replaced with the dot product

$$\boldsymbol{n}_j \cdot \boldsymbol{n} = \sin\theta_j \sin\theta_n \cos\varphi_n + \cos\theta_j \cos\theta_n. \tag{30}$$

### C. Arbitrary neutron wave packet

Let us now proceed to an arbitrary transverse wave function $\psi(\boldsymbol{p}_n)$ of the initial neutron. The polarization of each plane wave component again can generally depend on $\boldsymbol{p}_n$. The corresponding $S$-matrix element is

$$\mathcal{S}_{\text{packet}} = \int \frac{d^2 \boldsymbol{p}_n}{(2\pi)^2} \psi(\boldsymbol{p}_n)\mathcal{S}_{\text{PW}}(\boldsymbol{p}_n). \tag{31}$$

The square of the matrix element becomes

$$\begin{aligned}
|\mathcal{S}_{\text{packet}}|^2 &= \int \frac{d^2 \boldsymbol{p}_n}{(2\pi)^2}\frac{d^2 \boldsymbol{p}'_n}{(2\pi)^2} \psi(\boldsymbol{p}_n)\psi^*(\boldsymbol{p}'_n)\mathcal{S}_{\text{PW}}(\boldsymbol{p}_n)\mathcal{S}^*_{\text{PW}}(\boldsymbol{p}'_n)\\
&\propto \int \frac{d^2 \boldsymbol{p}_n}{(2\pi)^2}\frac{d^2 \boldsymbol{p}'_n}{(2\pi)^2} \psi(\boldsymbol{p}_n)\psi^*(\boldsymbol{p}'_n)\delta^{(2)}(\boldsymbol{p}_n - \boldsymbol{p}_p - \boldsymbol{p}_e - \boldsymbol{p}_{\bar{\nu}_e})\delta^{(2)}(\boldsymbol{p}'_n - \boldsymbol{p}_p - \boldsymbol{p}_e - \boldsymbol{p}_{\bar{\nu}_e})\mathcal{M}(\boldsymbol{p}_n)\mathcal{M}^*(\boldsymbol{p}'_n)\\
&= \int \frac{d^2 \boldsymbol{p}_n}{(2\pi)^4} \psi(\boldsymbol{p}_n)\psi^*(\boldsymbol{p}_n)\delta^{(2)}(\boldsymbol{p}_n - \boldsymbol{p}_p - \boldsymbol{p}_e - \boldsymbol{p}_{\bar{\nu}_e})|\mathcal{M}(\boldsymbol{p}_n)|^2\\
&= \int \frac{p_n d p_n d\varphi_n}{(2\pi)^4}|\psi(\boldsymbol{p}_n)|^2\delta^{(2)}(\boldsymbol{p}_n - \boldsymbol{p}_p - \boldsymbol{p}_e - \boldsymbol{p}_{\bar{\nu}_e})|\mathcal{M}(\boldsymbol{p}_n)|^2\\
&\propto \int \frac{d^2 \boldsymbol{p}_n}{(2\pi)^2}|\psi(\boldsymbol{p}_n)|^2|\mathcal{S}_{\text{PW}}(\boldsymbol{p}_n)|^2.
\end{aligned} \tag{32}$$

Thus, there appears no dependence on the phase of the wave packet in the decay rate. For an arbitrary wave function one can calculate the decay rate numerically by two-dimensional averaging of the plane-wave decay rate with the probability density of the packet. Meanwhile, one still needs to take into account the kinematic conditions of every single plane wave included in the superposition separately, which can be especially complicated if the polarization is a function of momentum.

### D. Laguerre-Gaussian and "spin-orbit" states

Although specifying the polarization of every plane wave in the wave packet seems a straightforward way to construct a quantum state with a nonuniform polarization field, it is not obvious how such states can be obtained experimentally. In this section we proceed with the neutron states with exotic polarizations, but within a slightly different formalism introduced in the works of Sarenac and colleagues [7,8,16]. The so-called "spin-orbit" states of neutrons, which the authors both described theoretically and obtained in experiments, are essentially the various superpositions of two homogeneously polarized Laguerre-Gaussian (LG) wave packets with different values of OAM and opposite spin directions. A few different methods such as magnetic spiral phase plates, quadrupole magnets, and magnetic prisms are proposed to prepare spin-orbit states of neutrons [7,16], where the last has been utilized in experiments [8].

The pure relativistic LG states for polarized fermions can be defined in momentum representation as

$$\langle \boldsymbol{p}|n,\ \ell,\ k_z,\ \boldsymbol{s}\rangle \equiv \psi_{n,\ell}(\boldsymbol{p})\frac{u(p,s)}{\sqrt{2E}}2\pi\delta(p_z - k_z), \tag{33}$$

where the transverse part of the wave function is

$$\psi_{n,\ell}(\boldsymbol{p}) = N_{\text{LG}}\frac{p_\perp^{|\ell|}}{\sigma^{|\ell|}}L_n^{|\ell|}\left(\frac{p_\perp^2}{\sigma^2}\right)\exp i\ell\varphi_p - \frac{p_\perp^2}{2\sigma^2}. \tag{34}$$

Here $n = 0, 1, 2, \ldots$ is the radial quantum number, $\ell = 0, \pm 1, \pm 2, \ldots$ is the azimuthal quantum number, $k_z$ is the momentum along $z$ direction, and $\sigma$ is the wave packet width in the momentum space. The bispinor $u(p, s)$ normalized as $u^\dagger(p, s) u(p, s) = 2E$ corresponds to a plane wave with the four-momentum $p^\mu = (E, \boldsymbol{p})$ and the spin four-vector $s^\mu$, which transforms into the polarization vector $\boldsymbol{s}$ in the rest frame [see Eqs. (19)–(22)]. The state can be normalized as

$$\int |\psi_{n,\ell}(\boldsymbol{p})|^2 \frac{d^2 p}{(2\pi)^2} = 1, \quad (35)$$

then the normalization constant is given by

$$N_{\text{LG}} = \sqrt{\frac{n!}{\pi (n + |\ell|)!}}. \quad (36)$$

As we have already mentioned, the spin-orbit state is a superposition of two LG modes with different values of OAM and polarization. The general form of such a state is

$$|\psi\rangle = \frac{1}{\sqrt{2}}(|n_1, \ell_1, k_z, \uparrow_z\rangle + e^{ib}|n_2, \ell_2, k_z, \downarrow_z\rangle), \quad (37)$$

where $b$ is an arbitrary relative phase. In the nonrelativistic limit spin and the spatial degree of freedom are decoupled, and this state can be thought of as entangled [7]:

$$|\psi\rangle_{\text{nonrel}} = \frac{1}{\sqrt{2}}(|n_1, \ell_1, k_z\rangle \otimes |\uparrow_z\rangle + e^{ib}|n_2, \ell_2, k_z\rangle \otimes |\downarrow_z\rangle). \quad (38)$$

The polarization of such state can be visualized by a coordinate-space spin field in the transverse plane of the wave packet. Denoting the nonrelativistic wave function of the LG state in position representation as

$$\phi_{n,\ell}(\boldsymbol{r}) = \int \psi_{n,\ell}(\boldsymbol{p}) e^{i\boldsymbol{p}\boldsymbol{r}} \frac{d^3 p}{(2\pi)^3} \quad (39)$$

and calculating the averages of nonrelativistic spin operator $\hat{\boldsymbol{s}} = 1/2 \hat{\boldsymbol{\sigma}}$, we obtain the inhomogeneous spin density in the polar coordinates $(r, \varphi)$:

$$\langle s \rangle_r = |\phi_{n_1,\ell_1}(r) \phi_{n_2,\ell_2}(r)| \cos[\varphi(\Delta \ell - 1) + b], \quad (40)$$

$$\langle s \rangle_\varphi = |\phi_{n_1,\ell_1}(r) \phi_{n_2,\ell_2}(r)| \sin[\varphi(\Delta \ell - 1) + b], \quad (41)$$

where the OAM difference $\Delta \ell = \ell_2 - \ell_1$. From this expression it is clearly seen that that the spin density distribution is rotationally invariant only for $\Delta \ell = 1$. A state with this specific ratio of OAM is nothing more than an eigenstate of total angular momentum projection onto the $z$ axis, which indeed has to be rotationally invariant. Generally speaking, for any other spin-orbit state with $\Delta \ell \neq 1$ its spin distribution is invariant under discrete rotations about $z$ axis at certain angles, i.e., integer multiplies of $2\pi/|\Delta \ell - 1|$. This kind of symmetry is simply described by the $C_{|\Delta \ell - 1|}$ (or $Z_{|\Delta \ell - 1|}$) symmetry group [18].

The corresponding $S$-matrix element for the decay of a neutron in such state is

$$\mathcal{S}_{\text{s-o}} = \frac{1}{\sqrt{2}} \int \frac{d^2 p_n}{(2\pi)^2} \{\psi_{n_1,\ell_1}(\boldsymbol{p}_n) \mathcal{S}_{\text{PW}}(\boldsymbol{p}_n, \boldsymbol{e}_z) + \psi_{n_2,\ell_2}(\boldsymbol{p}_n) \mathcal{S}_{\text{PW}}(\boldsymbol{p}_n, -\boldsymbol{e}_z)\}, \quad (42)$$

where we emphasize that the plane-wave matrix elements $\mathcal{S}_{\text{PW}}(\boldsymbol{p}, \boldsymbol{s})$ depend on the polarization by explicitly specifying the argument $\boldsymbol{s}$. Taking the square of the $S$ matrix yields two terms similar to previously derived ones along with the cross-term, which is more complicated to analyze:

$$|\mathcal{S}_{\text{s-o}}|^2 = \frac{1}{2} \int \frac{d^2 p_n}{(2\pi)^2} \{|\psi_{n_1,\ell_1}(\boldsymbol{p}_n)|^2 |\mathcal{S}_{\text{PW}}(\boldsymbol{p}_n, \boldsymbol{e}_z)|^2 + |\psi_{n_2,\ell_2}(\boldsymbol{p}_n)|^2 |\mathcal{S}_{\text{PW}}(\boldsymbol{p}_n, -\boldsymbol{e}_z)|^2 + 2\text{Re}[\psi_{n_1,\ell_1}(\boldsymbol{p}_n) \psi^*_{n_2,\ell_2}(\boldsymbol{p}_n) \mathcal{S}_{\text{PW}}(\boldsymbol{p}_n, \boldsymbol{e}_z) \times \mathcal{S}^*_{\text{PW}}(\boldsymbol{p}_n, -\boldsymbol{e}_z)]\}. \quad (43)$$

The product of $S$-matrix elements with different polarizations differs from the regular squared amplitude only by the nondiagonal element of the polarization density matrix, $u(p_n, s) \bar{u}(p_n, -s)$, the closed-form expression for which is rather cumbersome (see [29–33]):

$$u(p_n, \xi s) \bar{u}(p_n, -\xi s) = \tfrac{1}{2}(\slashed{p}_n + m_n) \gamma_5 (\slashed{n}_1 + \xi i \slashed{n}_2) \quad (44)$$

with $\xi = \pm 1$. The four-vectors $n_1^\mu$ and $n_2^\mu$ are defined by Lorentz boosts (from the neutron rest frame to the laboratory frame) of the vectors $\boldsymbol{\tau}_1$ and $\boldsymbol{\tau}_2$, which in turn are two orthogonal unit vectors that form a right-handed triad together with $\boldsymbol{s}$ in the three-dimensional space. These two vectors are defined up to a rotation around the direction of $\boldsymbol{s}$, which should be consistent with the phase of the bispinors $u(p_n, \pm \xi s)$ [31,34].

To construct the spin-orbit states we need only LG states with *homogeneous* polarization along the $z$ axis. Thus, the spin four-vector in the rest frame is $\boldsymbol{s} = \boldsymbol{e}_z$, and the auxiliary vectors become simply $\boldsymbol{\tau}_1 = \boldsymbol{e}_x$ and $\boldsymbol{\tau}_2 = \boldsymbol{e}_y$. The corresponding boost yields [cf. (22)]

$$n_1^\mu = (\gamma \beta \sin \theta_n \cos \varphi_n, \ \boldsymbol{e}_x + \sin \theta_n \cos \varphi_n (\gamma - 1) \boldsymbol{n}); \quad (45)$$

$$n_2^\mu = (\gamma \beta \sin \theta_n \sin \varphi_n, \ \boldsymbol{e}_y + \sin \theta_n \sin \varphi_n (\gamma - 1) \boldsymbol{n}). \quad (46)$$

It can be easily shown that the final expression for the cross-term $\mathcal{S}_{\text{PW}}(\boldsymbol{p}_n, s) \mathcal{S}^*_{\text{PW}}(\boldsymbol{p}_n, -s)$ (summed over polarizations of the final particles and integrated over the phase space) differs from the unpolarized decay rate (A3) only by the substitution [cf. the change (24)]

$$p_n^\mu \to m_n (n_1^\mu \pm i n_2^\mu). \quad (47)$$

Finally, note that spin-orbit states are not azimuthally invariant in contrast to the unpolarized vortex state and the polarization states introduced in [17]. Thus, the differential decay rate acquires the additional dependence on the azimuthal angle of the outgoing electron/proton. In this case the direction of the detected particle should be defined as

$$\boldsymbol{n}_j = (\sin \theta_j \cos \phi_j, \sin \theta_j \sin \phi_j, \cos \theta_j) \quad (48)$$

rather than setting $\phi_j$ to a fixed arbitrary value.

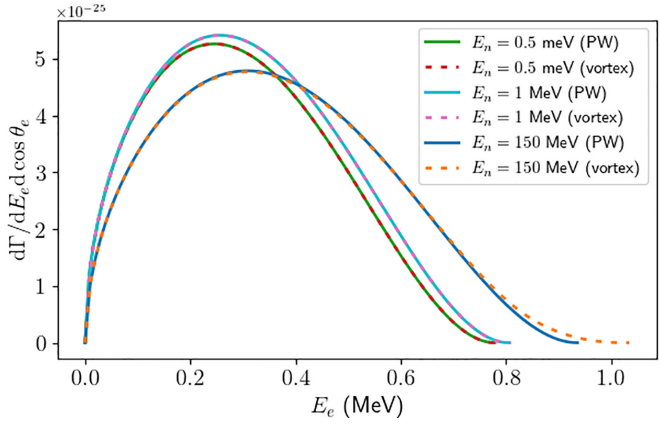

FIG. 2. Electron energy spectrum from unpolarized PW and Bessel ($\theta_n = 0.1$) neutron decays calculated at fixed $\theta_e = \pi/3$. The energies $E_e$, $E_p$, and $E_n$ here and below denote only kinetic energies. The spectra coincide with the results of [15] up to the factor of 2 (here the decay rate is twice smaller).

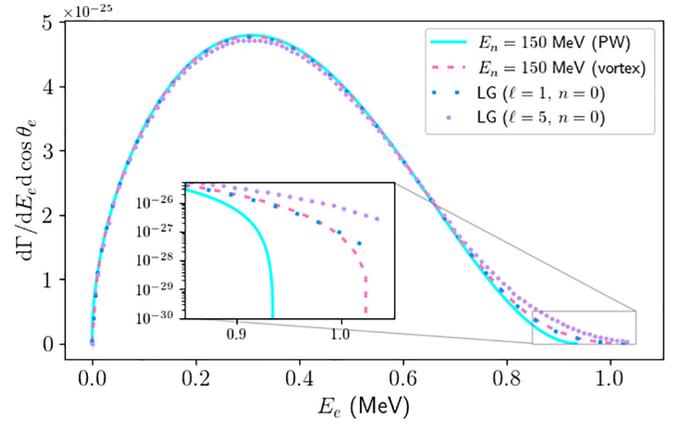

FIG. 3. Electron energy spectrum from 150 MeV unpolarized PW, Bessel ($\theta_n = 0.1$), and LG [$\sigma_p = \frac{2}{3} p_z \tan(\theta_n)$] neutron decays calculated at fixed $\theta_e = \pi/3$.

## IV. RESULTS AND DISCUSSION

In this section we present numerical results for the SAD of electrons and protons in beta-decay of vortex and spin-orbit neutron states. The choice of parameters in the graphs—energy, vortex cone opening angle, angle of observation, etc.—is mainly due to our desire to highlight the distortion of distributions compared to the plane-wave case. However, some values used in the following analysis should be justified. First, typical energies of neutrons in vortex and spin-orbit states available nowadays in the experiments are of the order of 1–10 meV [4,5,8]. Although methods of generation of structured neutrons with higher energies are not yet explored, there seem to be no fundamental difficulties in applying the quadruple magnet techniques to fast neutrons. Therefore, in this section we are guided by a much broader energy range (up to 300 MeV) dictated by the capabilities of existing sources of fast neutrons, such as China Spallation Neutron Source (CSNS) [24].

The differences of non-plane-wave neutron decay follow directly from the Eqs. (29), (32), and (43). According to these expressions, different plane-wave components of an arbitrary wave packet do not interfere in the decay, which is typical for processes with a single vortex initial state [17,18,28]. Thus, any expected deformations of SAD in comparison to the plane-wave case are due to the simultaneous contribution of several plane-waves to the decay.

In Fig. 2 we compare the electron energy spectra at fixed observation angle $\theta_e = \pi/3$ for plane-wave and Bessel neutrons decaying at different energies. The discrepancy between the spectra from vortex and plane-wave neutrons in this case turns out to be negligibly small and manifests itself only for extremely high energies of neutrons (∼150 MeV) and for $E_e \approx E_{e,\max}$.

Although the decay of Bessel neutrons is independent of the neutron's intrinsic OAM [15,17], this does *not* have to be the case for LG neutrons because the information about OAM is contained *not only in the phase but also in the spatial*

*profile of the LG packet*. To illustrate this effect, in Fig. 3 we duplicate the spectra corresponding to 150 MeV plane-wave and vortex neutrons from Fig. 2 and compare them to those for two different LG states: with $n = 0$, $\ell = 1$ and $n = 0$, $\ell = 5$. In order to match the Bessel and the "similar" LG neutron, we consider states with the same longitudinal momenta and set the width of the LG wave packet to be $\sigma_p = \frac{2}{3} p_z \tan(\theta_n)$. Since the LG packet is delocalized in the transverse plane unlike the Bessel wave (in momentum space), one can expect the "spreading" of the spectra and angular distributions. However, while the spectrum indeed changes with $\ell$, specifically broadening in the vicinity of the maximal energy, the modification is significant only in a narrow energy range even for such relativistic neutrons (see inset of Fig. 3).

The situation is different for SAD of the protons. Due to the proton large mass, which is almost equal to the neutron's, the spectra become much more sensitive to the structure of the neutron's wave packet and to its energy. In the proton energy spectra at fixed observation angle (Fig. 4) one can notice a distinct contrast between the plane-wave and vortex neutron decay already at energies of the order of 0.5 keV. The effect consists of the broadening of the spectrum analogously to the electron spectra. In Fig. 5 we again duplicate two spectra from Fig. 4 (corresponding to the neutron energy of 0.5 keV) and compare them with the spectra from LG neutrons. Here the broadening of the spectrum for larger values of $\ell$ becomes more pronounced than for the electron spectrum in Fig. 3.

One can expect another important kinematic effect in the angular distributions of the protons, which arises when the neutron's energy exceeds the value of ≈ 750 eV. This threshold indicates the existence of a maximal angle at which every plane-wave component can emit a proton. While for a plane-wave neutron the existence of a critical angle leads to the protons being emitted in the narrow forward cone, for Bessel neutrons the angular distribution of protons is expected to have a peak at the vortex cone opening angle $\theta_n$. This is demonstrated in Fig. 6, where we compare the angular distributions of protons from plane-wave neutrons, Bessel neutrons with different opening angles, and corresponding LG neutrons. The chosen energy of 75 keV corresponds to the critical angle $\xi_{\max} = 0.1$. For the opening angles $\theta_n \geqslant 0.1$ we indeed observe a forward dip, a peak at $\theta_p = \theta_n$, and a finite distribution width which equals $2\xi_{\max}$. For the matching LG neutrons the angular distributions are smeared out in such a manner that the decrease near zero becomes less distinct. Note that for the same parameters of the neutron the angular distribution of electrons does not exhibit these features and does not qualitatively distinguish between plane-wave and vortex neutrons (see Fig. 7). This is due to the fact that the effect of critical angle can be observed only for *unrealistically large* energy of the neutron ($E_n \gtrsim 1.5$ GeV), which we do not consider here.

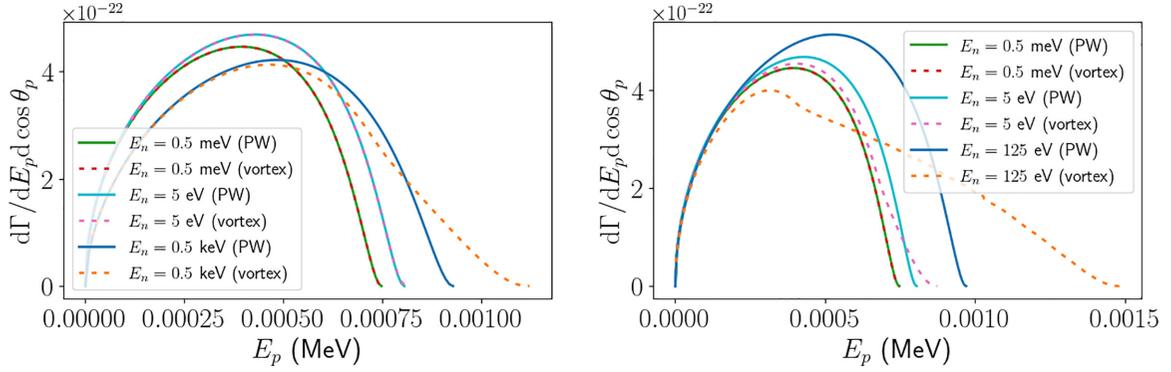

FIG. 4. Proton energy spectra from unpolarized PW and Bessel [for $\theta_n = 0.1$ (left) and unrealistic $\theta_n = 1$ (right)] neutron decays calculated at fixed $\theta_p = \pi/3$.

Finally, we discuss the decay of neutrons in exotic polarization states. The nonuniform spin field described by Eq. (40) becomes rotationally invariant when $\Delta\ell = 1$, while in all other cases the spin density of the neutron wave packet depends on the azimuthal angle. As a consequence, one can expect the angular distributions of the emitted particles to be *azimuthally dependent* as well. We have already discussed that distortions of the SAD can generally be observed for protons at much lower neutron energies than for electrons, therefore here we limit ourselves to the protons. Since it is only spin-orbit states of neutrons with $|\ell_2 - \ell_1| = 1$ that have been recently obtained in experiments [8], we mainly focus on this case in the examples.

First, in Fig. 8 we compare the azimuthal dependencies of the angular distributions at fixed polar angles $\theta_p$ for three different values of neutron energy: 10 meV, 0.5 keV, and 4.5 keV. For convenience the distributions are normalized by the values at their minimum. Since in this example the OAM difference $\Delta\ell = -1$, the angular distributions reveal the same rotational symmetry $C_2$ as the polarization state of the decaying neutron [18]. For cold neutrons the maximal azimuthal variation is of the order of $5 \times 10^{-4}$ at the observation angle $\theta_p = \pi/2$. For faster although still nonrelativistic neutrons, the asymmetry reaches the maximal value around 0.2 at smaller polar angle $\theta_p \approx 1$ rad. For the energy of 4.5 keV, which already enables the existence of a critical angle about 0.4 rad, the variation of the azimuthal distribution reaches the value of 5.

If the relative phase $b$ of two LG modes is not equal to zero, the two-peak pattern is simply shifted along the azimuthal direction (rotated around the $z$ axis), as shown in Fig. 9. For $b = \pi/2$ and $b = \pi$ the minima are located at $\varphi_p = \pi/4$, $\varphi_p = 5\pi/4$ and $\varphi_p = 0$, $\varphi_p = \pi$, respectively, while for $b = 0$ (see Fig. 8) the minimum points are $\varphi_p = \pi/2$, $\varphi_p = 3\pi/2$. This effect is also explained by the symmetry of the spin density

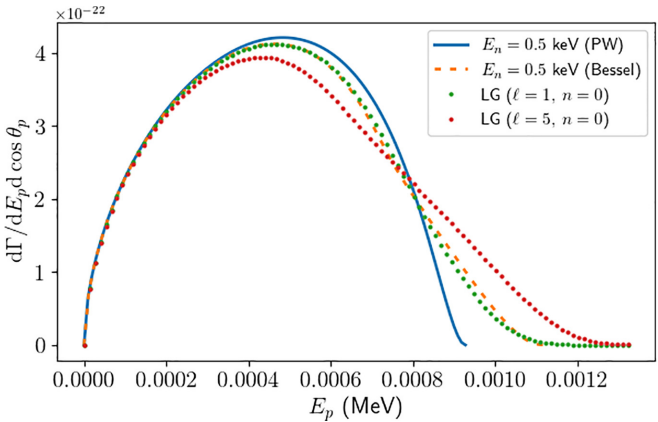

FIG. 5. Proton energy spectra from 0.5 keV unpolarized PW, Bessel ($\theta_n = 0.1$), and LG [$\sigma_p = \frac{2}{3} p_z \tan(\theta_n)$] neutron decays calculated at fixed $\theta_p = \pi/3$.

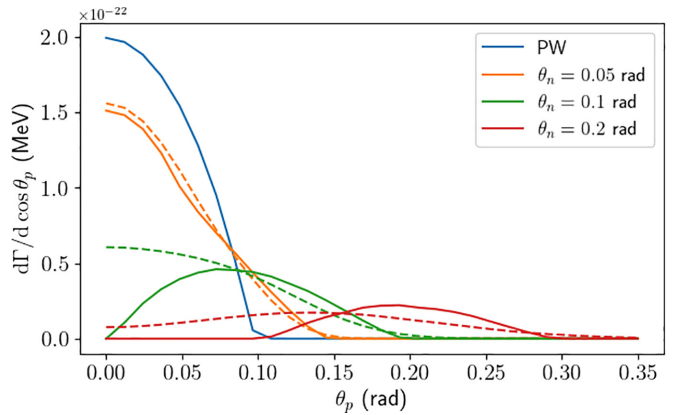

FIG. 6. Angular distribution of protons from unpolarized 75 keV PW, Bessel (solid lines), and corresponding LG neutrons with $n = 0$, $\ell = 1$ (dashed lines).

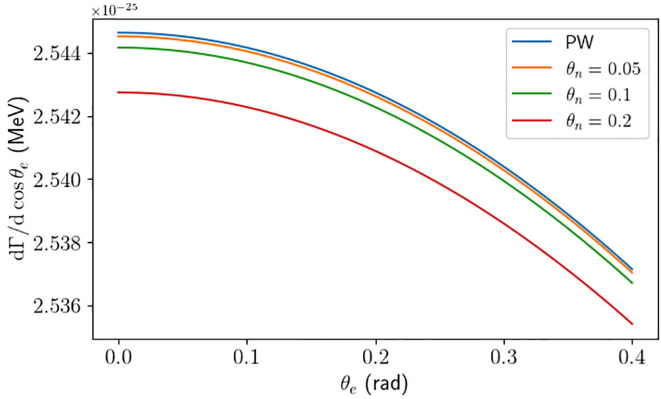

FIG. 7. Electron angular distribution from 75 keV unpolarized PW and Bessel neutrons.

(40) because the phase $b$ is essentially the angle at which the spin field is "rotated" simultaneously at every point.

Although the nonzero value of the radial quantum number $n$ does not qualitatively change the azimuthal behavior of the distribution, it slightly increases the asymmetry, as shown in Fig. 10 (cf. Fig. 8). The rotational symmetry of higher orders is demonstrated in Fig. 11: for $\Delta\ell = -2$ and $\Delta\ell = -3$ we obtain three and four peaks in the distribution, respectively. The amplitude of azimuthal variation, however, noticeably decreases with the growth of $|\Delta\ell|$.

## V. CONCLUSION

In the current work we have developed a formalism for beta decay of a neutron in an arbitrary non-plane-wave (structured) state. The focus has been put on three examples, motivated by the existing experiments [4–6,8]: unpolarized vortex (Bessel) neutrons carrying nonzero OAM, LG packets, and spin-correlated OAM (spin-orbit) states, which exhibit an inhomogeneous polarization.

The differences between the decays of structured and plane-wave neutrons are solely due to kinematics, because the decay rate is an incoherent superposition of the plane-wave contributions. We have shown that the SAD of protons generally displays much higher sensitivity to the non-plane-wave structure of the initial packet than that of the electrons. One of the important effects responsible for the modifications of angular distributions is a limit on maximal angle between the plane-wave proton and neutron directions, which can be observed for neutrons with kinetic energy higher than 750 eV. Due to the conical momentum distribution of the Bessel state, this leads to the protons' distribution having a peak at the opening angle of the cone and a forward minimum. The similar effect exists in many other processes involving vortex states [17,35–37].

Due to the absence of interference between plane waves constituting the Bessel state, the decay rate becomes OAM independent [17,28]. However, for LG states of the neutron it does depend on the OAM because not only the phase of the LG state but also its probability density are OAM dependent.

As for the spin-orbit states, which are nothing more than a superposition of LG states with opposite polarizations, our key finding is that the angular distribution acquires the azimuthal dependence by inheriting the discrete rotational symmetry from the nonuniform spin density of the packet (a similar feature was recently predicted in muon decay [18]). In contrast to the maximal proton deflection angle, this effect can possibly be observed even at currently available energies of structured neutron beams ($\sim$10 meV).

## ACKNOWLEDGMENTS

We are grateful to D. Grosman and G. Sizykh for many fruitful discussions and advice. The studies in Sec. III A are supported by the Government of the Russian Federation through the ITMO Fellowship and Professorship Program. The studies in Secs. III B and III C are supported by the Foundation for the Advancement of Theoretical Physics and Mathematics "BASIS". The studies in Sec. III D are supported by the Russian Science Foundation (Project No. 23-62-10026) [38].

## DATA AVAILABILITY

No data were created or analyzed in this study.

## APPENDIX A: PHASE SPACE INTEGRAL

We start by considering the integral

$$I^{\alpha\beta} = \int p_i^\alpha p_{\bar{\nu}_e}^\beta \frac{d^3 p_i}{\omega_i} \frac{d^3 p_{\bar{\nu}_e}}{\omega_{\bar{\nu}_e}} \delta^4(p_i + p_{\bar{\nu}_e} - q), \quad (A1)$$

where $\omega_i = \sqrt{m_i^2 + \boldsymbol{p}_i^2}$, $\omega_{\bar{\nu}_e} = |\boldsymbol{p}_{\bar{\nu}_e}|$. Importantly, this a Lorentz-covariant object, which can depend only on $q^2$ and $m_i$. This allows us to guess the answer as the sum of two mutually orthogonal terms:

$$I^{\alpha\beta} = A(q^2, m_i)(q^2 g^{\alpha\beta} + 2q^\alpha q^\beta) \\ + B(q^2, m_i)(q^2 g^{\alpha\beta} - 2q^\alpha q^\beta), \quad (A2)$$

where $q = p_{\bar{\nu}_e} + p_i = p_n - p_j = (\omega, \boldsymbol{q})$. The differential decay rate hence becomes

$$\frac{d\Gamma_{PW}}{dE_j d\Omega_j} \propto A(q, m_i)[q^2(p_n p_j) + 2(qp_n)(qp_j)] \\ + B(q, m_i)[q^2(p_n p_j) - 2(qp_n)(qp_j)]. \quad (A3)$$

In the case of massless final particles, which is discussed in [25] for a similar process of muon decay, $A$ and $B$ are constants, but it is not possible when the electron and proton are treated as massive. The coefficients are calculated as follows:

$$I^{\alpha\beta}(q^2 g^{\alpha\beta} - 2q^\alpha q^\beta) \\ = 4Bq^4 = \int \left[(p_{\bar{\nu}_e} p_i)q^2 - 2(qp_{\bar{\nu}_e})(qp_i)\right] \frac{d^3 p_i}{\omega_i} \frac{d^3 p_{\bar{\nu}_e}}{\omega_{\bar{\nu}_e}}. \quad (A4)$$

Here $q^2 = (p_{\bar{\nu}_e} + p_i)^2 = m_i^2 + 2(p_{\bar{\nu}_e} p_i)$, $qp_{\bar{\nu}_e} = p_i p_{\bar{\nu}_e}$, $qp_i = p_{\bar{\nu}_e} p_i + m_i^2$. Thus,

$$4Bq^4 = \frac{m_i^2}{2}(m_i^2 - q^2) \int \frac{d^3 p_i}{\omega_i} \frac{d^3 p_{\bar{\nu}_e}}{\omega_{\bar{\nu}_e}} \delta^4(p_i + p_{\bar{\nu}_e} - q). \quad (A5)$$

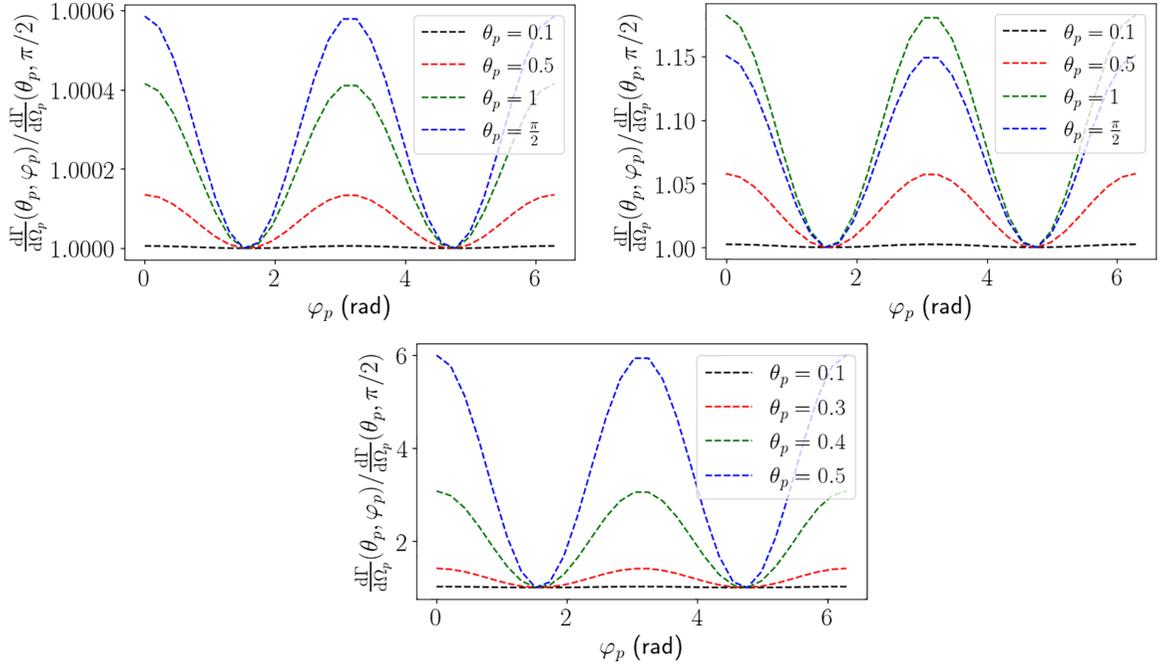

FIG. 8. Azimuthal distribution of protons calculated at different angles $\theta_p$ from decay of 10 meV (left), 0.5 keV (middle), and 4.5 keV (right) spin-orbit neutron states ($n_1 = n_2 = 0$, $\ell_1 = 0$, $\ell_2 = -1$, $b = 0$).

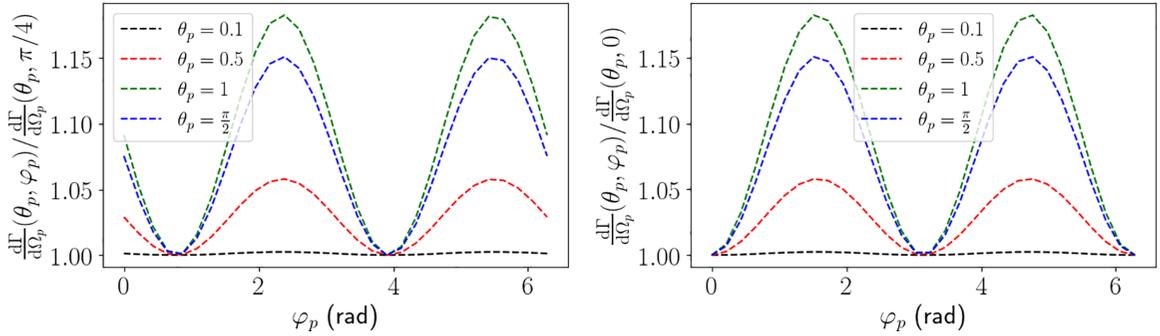

FIG. 9. Azimuthal distribution of protons from decay of 0.5 keV spin-orbit neutron states ($n_1 = n_2 = 0$, $\ell_1 = 0$, $\ell_2 = -1$) for $b = \pi/2$ (left) and $b = \pi$ (right).

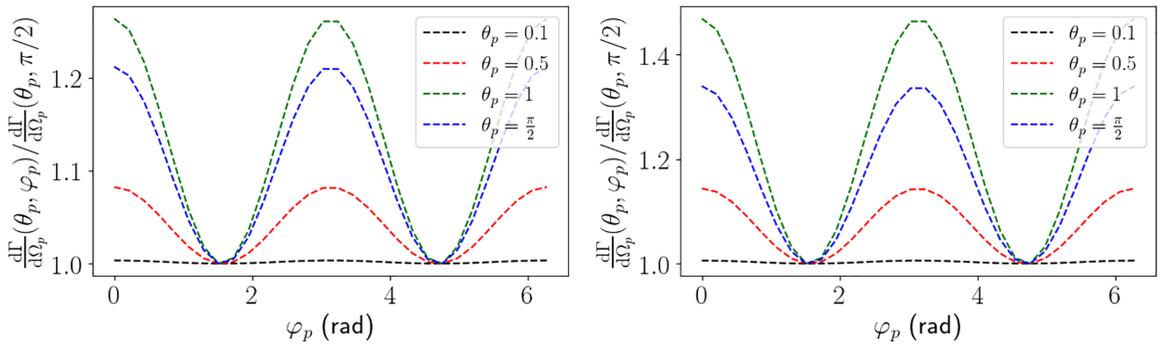

FIG. 10. Azimuthal distribution of protons from decay of 0.5 keV spin-orbit neutron state ($\ell_1 = 0$, $\ell_2 = -1$, $b = 0$) for $n_1 = n_2 = 1$ (left) and $n_1 = n_2 = 5$ (right).

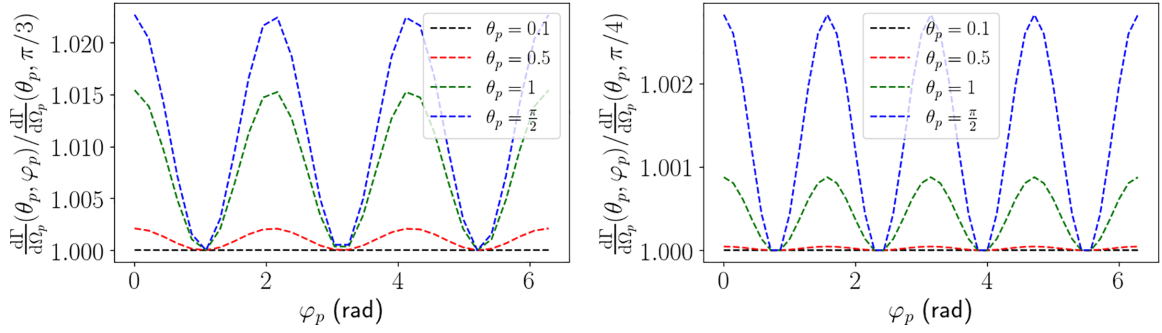

FIG. 11. Azimuthal distribution of protons from decay of 0.5 keV spin-orbit neutron state ($n_1 = n_2 = 0$, $\ell_1 = 0$, $b = 0$) for $\ell_2 = -2$ (left) and $\ell_2 = -3$ (right).

The remaining integral is Lorentz invariant and can be evaluated in any reference frame. This is most easily done in the center-of-mass frame of the $i$-particle and neutrino, where $q = (\omega, 0)$:

$$\int \frac{d^3 p_i}{\sqrt{m_i^2 + \boldsymbol{p}_i^2}} \frac{d^3 p_{\bar{\nu}_e}}{|\boldsymbol{p}_{\bar{\nu}_e}|} \delta^4(p_i + p_{\bar{\nu}_e} - q)$$

$$= \int \frac{d^3 p_i}{|\boldsymbol{p}_i|\sqrt{m_i^2 + \boldsymbol{p}_i^2}} \delta(\sqrt{m_i^2 + \boldsymbol{p}_i^2} + |\boldsymbol{p}_i| - \omega)$$

$$= 4\pi \int \frac{|\boldsymbol{p}_i| d|\boldsymbol{p}_i|}{\sqrt{m_i^2 + \boldsymbol{p}_i^2}} \delta(\sqrt{m_i^2 + \boldsymbol{p}_i^2} + |\boldsymbol{p}_i| - \omega)$$

$$= 4\pi \int \frac{|\boldsymbol{p}_i| d|\boldsymbol{p}_i|}{\sqrt{m_i^2 + \boldsymbol{p}_i^2}} \frac{\omega^2 + m_i^2}{2\omega^2} \delta\left(|\boldsymbol{p}_i| - \frac{\omega}{2} + \frac{m_i^2}{2\omega}\right)$$

$$= \frac{\frac{\omega}{2} - \frac{m_i^2}{2\omega}}{\frac{\omega}{2} + \frac{m_i^2}{2\omega}} 2\pi \left(1 + \frac{m_i^2}{\omega^2}\right)$$

$$= 2\pi \left(1 - \frac{m_i^2}{q^2}\right). \tag{A6}$$

Thus, we obtain the coefficient

$$B(q, m_i) = -\frac{\pi m_i^2}{4 q^2}\left(1 - \frac{m_i^2}{q^2}\right)^2. \tag{A7}$$

An analogous calculation for the coefficient $A(q, m_i)$ yields

$$A(q, m_i) = \frac{\pi}{6}\left(1 - \frac{m_i^2}{q^2}\right)\left(1 - \frac{m_i^2}{2 q^2} - \frac{m_i^4}{2 q^4}\right). \tag{A8}$$

Now, rewriting the differential of the remaining particle phase volume as

$$d^3 p_j = |\boldsymbol{p}_j|^2 d|\boldsymbol{p}_j| d\Omega_j = |\boldsymbol{p}_j| E_j dE_j d\Omega_j$$

$$= \sqrt{E_j^2 - m_j^2} E_j dE_j d\Omega_j \tag{A9}$$

and using the relations

$$(q p_n) = m_n^2 - (p_n p_j), \quad (q p_j) = (p_n p_j) - m_j^2, \tag{A10}$$

we finally obtain the decay rate (16).

Using the substitution (24), for the decay rate of a polarized neutron we find

$$\frac{d\Gamma}{dE_j d\Omega_j} \propto A(q, m_i)\{[q^2(p_n p_j) + 2(q p_n)(q p_j)]$$

$$- m_n[q^2(s p_j) + 2(q s)(q p_j)]\}$$

$$+ B(q, m_i)\{[q^2(p_n p_j) - 2(q p_n)(q p_j)]$$

$$- m_n[q^2(s p_j) - 2(q s)(q p_j)]\}. \tag{A11}$$

Here the dot products involving the spin four-vector are

$$(s p_j) = E_j \gamma \beta (\boldsymbol{s} \cdot \boldsymbol{n}) - \sqrt{E_j^2 - m_j^2}$$
$$\times (\boldsymbol{s} \cdot \boldsymbol{n}_j + (\gamma - 1)(\boldsymbol{n}_j \cdot \boldsymbol{n})(\boldsymbol{s} \cdot \boldsymbol{n})) \tag{A12}$$

and

$$(s q) = (s p_n) - (s p_j)$$
$$= \gamma \beta (\boldsymbol{s} \cdot \boldsymbol{n})\sqrt{m_n^2 + \boldsymbol{p}_n^2} - (\boldsymbol{s} \cdot \boldsymbol{n})|\boldsymbol{p}_n|$$
$$- (\gamma - 1)(\boldsymbol{s} \cdot \boldsymbol{n})^2 |\boldsymbol{p}_n| - (s p_j). \tag{A13}$$

## APPENDIX B: BESSEL NEUTRON DECAY

The square of the twisted matrix element is expressed as

$$|S_{\text{tw}}|^2 = \int \frac{d^2 \boldsymbol{p}_n}{(2\pi)^2} \frac{d^2 \boldsymbol{p}'_n}{(2\pi)^2} a_{\kappa\ell}(\boldsymbol{p}_n) a^*_{\kappa\ell}(\boldsymbol{p}'_n) S_{\text{PW}}(\boldsymbol{p}_n) S^*_{\text{PW}}(\boldsymbol{p}'_n)$$

$$\propto \int \frac{d^2 \boldsymbol{p}_n}{(2\pi)^2} \frac{d^2 \boldsymbol{p}'_n}{(2\pi)^2} a_{\kappa\ell}(\boldsymbol{p}_n) a^*_{\kappa\ell}(\boldsymbol{p}'_n)$$

$$\times \delta^{(2)}(\boldsymbol{p}_n - \boldsymbol{p}_p - \boldsymbol{p}_e - \boldsymbol{p}_{\bar{\nu}_e})$$

$$\times \delta^{(2)}(\boldsymbol{p}'_n - \boldsymbol{p}_p - \boldsymbol{p}_e - \boldsymbol{p}_{\bar{\nu}_e}) \mathcal{M}(\boldsymbol{p}_n)\mathcal{M}^*(\boldsymbol{p}'_n)$$

$$= \int \frac{d^2 \boldsymbol{p}_n}{(2\pi)^4} a_{\kappa\ell}(\boldsymbol{p}_n) a^*_{\kappa\ell}(\boldsymbol{p}_n)$$

$$\times \delta^{(2)}(\boldsymbol{p}_n - \boldsymbol{p}_p - \boldsymbol{p}_e - \boldsymbol{p}_{\bar{\nu}_e})|\mathcal{M}(\boldsymbol{p}_n)|^2$$

$$= \int \frac{d\varphi_n}{(2\pi)^3} \delta^{(2)}(\boldsymbol{p}_n - \boldsymbol{p}_p - \boldsymbol{p}_e - \boldsymbol{p}_{\bar{\nu}_e})|\mathcal{M}(\boldsymbol{p}_n)|^2. \tag{B1}$$

The square of $a_{\kappa\ell}(\boldsymbol{p}_n)$ contains a radial delta-function squared, which is regularized as

$$[\delta(|\boldsymbol{p}_\perp|-\kappa)]^2 = \delta(|\boldsymbol{p}_\perp|-\kappa)\delta(0) \to \delta(|\boldsymbol{p}_\perp|-\kappa)\frac{R}{\pi}, \quad \text{(B2)}$$

since one can treat the delta function as

$$\delta(0) = \int_0^\infty r\,\mathrm{d}r[J_\ell(\kappa r)]^2 \to \int_0^R r\,\mathrm{d}r[J_\ell(\kappa r)]^2 \approx \frac{R}{\pi}. \quad \text{(B3)}$$

(see [28,39]). The normalization of the twisted states also differs from the plane waves. To renormalize a plane wave to one particle per the entire volume $V$, the plane wave should be multiplied by such $N_{\mathrm{PW}}$ that

$$N_{\mathrm{PW}}^2 = \frac{1}{2EV},$$

while for a twisted state the normalization factor equals

$$N_{\mathrm{tw}}^2 = \frac{1}{2E}\frac{\pi}{RL},$$

where $L$ and $R$ are the length and radius of the normalization cylinder.

The twisted decay rate therefore becomes

$$\begin{aligned}
\mathrm{d}\Gamma_{\mathrm{tw}} &= \frac{1}{T}\frac{\pi}{2E_n RL}|\mathcal{S}_{\mathrm{tw}}|^2 \left(\frac{V\mathrm{d}^3 p_e}{(2\pi)^3 2E_e V}\right) \\
&\quad \times \left(\frac{V\mathrm{d}^3 p_p}{(2\pi)^3 2E_p V}\right)\left(\frac{V\mathrm{d}^3 p_{\bar{\nu}_e}}{(2\pi)^3 2E_\nu V}\right) \\
&= \frac{1}{2E_n}(2\pi)^4\delta(E_i-E_f)\delta(p_{zi}-p_{zf})\int\frac{\mathrm{d}\varphi_n}{2\pi}|\mathcal{M}(\boldsymbol{p}_n)|^2 \\
&\quad \times \delta^{(2)}(\boldsymbol{p}_n-\boldsymbol{p}_p-\boldsymbol{p}_e-\boldsymbol{p}_{\bar{\nu}_e}) \\
&\quad \times \left(\frac{\mathrm{d}^3 p_e}{(2\pi)^3 2E_e}\right)\left(\frac{\mathrm{d}^3 p_p}{(2\pi)^3 2E_p}\right)\left(\frac{\mathrm{d}^3 p_{\bar{\nu}_e}}{(2\pi)^3 2E_\nu}\right), \quad \text{(B4)}
\end{aligned}$$

where indices $i$ and $f$ mean initial and final (energy or momentum), respectively. Thus, the decay width of a Bessel vortex neutron is reduced to the azimuthal average of the plane-wave decay width.